\documentclass[smallextended]{svjour3}

\smartqed

\usepackage[pdftex]{graphicx}
\graphicspath{{./iir/}}
\DeclareGraphicsExtensions{.pdf}

\usepackage[cmex10]{amsmath}
\usepackage{amssymb}
\usepackage{relsize}
\usepackage{algorithmic, algorithm, listings}
\usepackage{mathptmx}
\usepackage{natbib}

\RequirePackage{amsmath, amssymb, relsize, fix-cm}

\newcommand{\bmdef}{{\buildrel{{\mathsmaller{\triangle}}}\over{=}}}
\newcommand{\bmvec}[1]{{\mathbf{#1}}}
\newcommand{\bmmat}[1]{{\mathbf{#1}}}

\newcommand{\bmcardinality}[1]{{\left\lvert{#1}\right\rvert}}

\newcommand{\bmsequence}[1]{{\left\langle{#1}\right\rangle}}

\newcommand{\bmindex}[1]{{\left[{#1}\right]}}

\newcommand{\bmtranspose}{{T}}

\journalname{GeNeDis}

\title{Blockchain For Mobile Health Applications}
\subtitle{Acceleration With GPU Computing}
\titlerunning{Blockchain For Mobile Health Applications}
\author{%
    Georgios Drakopoulos  \and
    Michail Marountas  \and
    Xenophon Liapakis  \and
    Giannis Tzimas  \and
    Phivos Mylonas  \and
    Spyros Sioutas
}%
\authorrunning{Drakopoulos et al.}
\institute{%
    G. Drakopoulos and P. Mylonas \at
    Department of Informatics, Ionian University, Greece\\
    \email{\{c16drak, fmylonas\}@ionio.gr}
\and
    X. Liapakis \at
    Interamerican SA, Greece\\
    \email{liapakisx@interamerican.gr}
\and
    G. Tzimas  \at
    Technological and Educational Institute of Western Greece, Antirrio Campus, Greece\\
    \email{tzimas@teimes.gr}
\and
    M. Marountas and S. Sioutas  \at
    Computer Engineering and Informatics Department, University of Patras, Greece\\
    \email{\{marounta, sioutas\}@ceid.upatras.gr}
}%

\date{Received: date / Accepted: date}

\def\bmwork{chapter}
\def\bmworkk{work}

\begin{document}

\maketitle

\begin{abstract}
Blockchain is a linearly linked, distributed, and very robust data structure. Originally proposed as part of the Bitcoin distributed stack, it found a number of applications in a number of fields, most notably in smart contracts, social media, secure IoT, and cryptocurrency mining. It ensures data integrity by distributing strongly encrypted data in widely redundant segments. Each new insertion requires verification and approval by the majority of the users of the blockchain. Both encryption and verification are computationally intensive tasks which cannot be solved with ordinary off-the-shelf CPUs. This has resulted in a renewed scientific interest in secure distributed communication and coordination protocols. Mobile health applications are growing progressively popular and have the enormous advantage of timely diagnosis of certain conditions. However, privacy concerns have been raised as mobile health application by default have access to highly sensitive personal data. This \bmwork{} presents concisely how blockchain can be applied to mobile health applications in order to enhance privacy.
\end{abstract}

\section{Introduction}\label{sec:intro}

\paragraph{}  Perhaps the most well studied recent advent in the domain of distributed comouting and data structures is that of blockchain. The latter acts as a public or private ledger and from a structural perspective is a linear, distributed, and robust data structure in the sense that not only the insertion of new data requires special permission from its stakeholders, mostly but not necessarily ordinary netizens with a legitimate vested interest in a given blockchain, which is obtained from specially designed consensus protocols, but also the true netizen identities participating to a given block chain as well as data contained therein are strongly encrypted, typically with a public key scheme such as SHA-256. Additionally, the exact location of data insertion is decided on the basis of a secure hash function. Finally, when the number of netizens participating to a blockchain is large, typically in the thousands, it becomes difficult to hack or game it as any malicious changes become visible almost immediately.

\paragraph{}  Having the computational properties just described, a blockchain is an excellent data structure for securely storing large volumes of information for a very broad spectrum of purposes including but not limited to smart contracts, digital health information, smart city and smart infrastructure status, financial macrotransactions as well as gaming and social media microtransactions, and insurance information. In fact, the blockchain as a data structure was initially part of the Bitcoin stack as described in \cite{nakamoto:2008} or as later explored in cryptocurrency surveys as for instance \cite{antonopoulos:2014} or \cite{antonopoulos:2017}. Since then, however it took a life of its own with numerous parties developing some version of the original blockchain for their own purposes.

\paragraph{}  Blockchain is not the only recent computationally intensive development. Fields like numerical and distributed deep learning such as the training of multilyayer convolutional and recurrent neural networks, complex systems simulation such as brain networks and protein-to-protein interaction networks, as large scale social network analysis are notorious for their quick scaling. One response to the need for additional computational power was the development of hardware aiming at massive parallelism through special purpose GPUs along with the associated software which can take advantage of such specialized hardware and can orchestrate the appopriate sequence of computations to derive the desired result. Google TensorFlow, a low level framework whose primary unit is a tensor as explained among others in \cite{abadi:2016:osdi}, namely a multidimensional array, belongs to this category.  

\paragraph{}  The primary objective of this \bmwork{} is to concentrate and succintly present the ways TensorFlow and GPU computation in conjunction with blockchain can empower applications in the domains of digital health and insurance market. As a secondary objective, the computational capabilities and the dataflow model of TensorFlow are analysed. 

\paragraph{}  The remaining of this \bmworkk{} is structured as follows. In section \ref{sec:work} the relevant scientific literature regarding blockchain, GPU computing, and their applications is briefly reviewed. The properties of the blockchain as well as these of TensorFlow are described in section \ref{sec:blockchain}, whereas the blockchain applications in the domains of digital health and insurance are explored in section \ref{sec:app}. The main findings of this \bmwork{} as well as possible future research directions are stated in section \ref{sec:more}. Finally, table \ref{tab:notation} summarises the notation of this \bmworkk.
\begin{table}
\caption{Notation of this \bmwork.}\label{tab:notation}
\begin{tabular}{ll}
    \hline\noalign{\smallskip}
    symbol  &  meaning  \cr
    \noalign{\smallskip}\hline\noalign{\smallskip}
    $\bmdef$            &  Definition or equality by definition    \cr
    $\bmsequence{s_k}$  &  Sequence with elements $s_k$            \cr
    $\bmcardinality{\bmsequence{s_k}}$  &  Sequence cardinality    \cr
    \noalign{\smallskip}\hline
\end{tabular}
\end{table}

\section{Previous Work}\label{sec:work}

\paragraph{}  Blockchains were formally introduced in the seminal Bitcoin work of \cite{nakamoto:2008}. Their technological innovation and the potential to become a disruptive technology was explored among others in \cite{barber:2012} and in \cite{cachin:2016}. The combination of blockchains with the IoT and their applications to the mainstream industrial sector in conjunction with the upcoming digital transformations of Industry 4.0 are the focus of \cite{miller:2018}. Practical ways and the associated challenges to implement a blockchain over IoT and edge computing are shown in \cite{zyskind:2015}. The financial prospects of Bitcoin in terms of wealth accumulation as well as the properties of Bitcoin versus the traditional fiat currency are the focus of a number of works, for instance \cite{antonopoulos:2014}, \cite{antonopoulos:2017}, \cite{kosba:2016}, \cite{swan:2015}, and \cite{bohme:2015}. The distributed implementation of blockchains is discussed in \cite{abbas:2018} and in \cite{pass:2017}, whereas security aspects of the blockchains are treated in \cite{puthal:2018}. A large number of blockchains besides the Bitcoin stack can be found in \cite{underwood:2016}.

\paragraph{}  Since the original public description of TensorFlow in \cite{abadi:2016:osdi} and in \cite{abadi:2016:sigplan} it was widely adopted from the deep learning community. In \cite{matthews:2017} a Gaussian process generator implemented with rudimentary TensorFlow operations is described in detail. For a new graph resilience metric based on paths see \cite{drakop:2018:ictai} along the lines of the regularization methodology of \cite{drakop:2017:webist}. The advantages of and the ways for visualising the TensorFlow computations are \cite{wongsuphasawat:2018}. For tensor applications in social network analysis such as multiway digital influence estimation see \cite{drakop:2017:snam}, community structure discovery \cite{drakop:2018:evos:neo4j}, and graph based k-means initialization \cite{drakop:2016:ijait}. Finally, for a genetic algorithm for clustering tensors containing linguistic and spatial data see \cite{drakop:2019:evos:tensor}.

\section{Parallelism and Blockchain}\label{sec:blockchain}

\subsection{Blockchains}

\paragraph{}  As their collective name suggests, from a structural point of view blockchains are, typically very long, linearly linked nodes. Each such node contains part of the postmarked information stored in the data structure along with some administrative information. The data stored in a blockchain can never be erased, although it can be updated provided all interested parties agree on that. Thus, both the original and the updated data are stored, making audits efficient. 

\paragraph{}  Perhaps the most important advantages of selecting a blockchain scheme besides the increased security are the following:
\begin{itemize}

    \item  Blockchains support a very large volume of transactions which can take place almost simultaneously because of their very inherent distributed nature. Therefore, their stakeholders can perform any desired number of transactions within a very reasonable amount of time without worrying about the exact transaction execution time, which in certain cases may influence the transaction cost.

    \item  The stakeholders of a given blockchain can stay informed of the global status of the blockchain in almost real time. Thus, not only can they perform transactions but they can also know their results almost immediately or at least at the moment the latter are actually executed. 

    \item  Blockchains, either public or private, offer full transparency since every participant to a given blockchain is free to validate any tranaction which took place within that blockchain. Additionally, the verification protocols are deliberately built so that verification be easy even for netizens with low computational resources, for instance a smartphone or a tablet. This reinforces the trust toward properly implemented and managed blockchains.

    \item  In the case of a catastrophic loss, a properly implemented blockchain can at least partially rebuild itself from the segments stored at the computers of its stakeholders. This is feasible given the increased redundancy integrated into a blockchain.

    \item  From a software engineering viewpoint, each blockchain node is a relatively simple construct and, therefore, it can be managed with little or no human intervention. Thus, a blockchain administrator is only required to control certain a few key parts of the data structure, making blockchains easy and inexpensive to maintain.  

    \item  Last but not least, any third parties and intermediaries are no longer necessary. The interested parties can directly communicate and get current quotes or any other vital pieces of information from each other.

\end{itemize}

\paragraph{}  Notice that blockchains are not immune to various sophisticated attacks, although the latter typically require considerable resources which are nowadays well within the capabilities of a dedicated hacker group or of a government agency. Although directly attacking the encryption protocols may not be a wise course of action, unless some knowledge of the private key is available, using a zero day exploit is.

\paragraph{}  As with any new technology, blockchain management software is by no means error free. However, most known attacks so far take on a completely different approach akin to a brute force attack by relying on big botnet networks in order to take charge of a small or medium sized blockchain. 

\paragraph{}  Yet another method, holistic in nature, for attacking a blockchain is through the use of control theory concepts. The current state, in any way that is estimated by the attacker, of a large blockchain is represented as a control vector $\bmvec{x}\bmindex{n}$. Then a usually linear state space model is formulated as follows:
\begin{align}
\bmvec{x}\bmindex{n+1} &  \:\bmdef\:
    \bmmat{A}\bmvec{x}\bmindex{n} + \bmvec{b} u\bmindex{n}  \cr
\bmvec{y}\bmindex{n+1} &  \:\bmdef\:
    \bmvec{c}^\bmtranspose \bmvec{x}\bmindex{n+1} + d u\bmindex{n}
\end{align}
If the attacker can insert an appropriate input sequence $\bmsequence{u_k}$, then, depending on the modelling correctness, he may bring the entire system to an undersirable state. Of course, such a sequence may not exist or its cardinality $\bmcardinality{\bmsequence{u_k}}$ might approach infinity.

\subsection{TensorFlow}

\paragraph{}  Google TensorFlow is a low level programming framework based on the dataflow programming paradigm and using tensors, namely multidimensional arrays, as its primary data structure. Originally developed for simulating brain networks, it is a powerful tool for deep learning. It has official APIs for Python and C++, whereas unofficial APIs are being developed for a number of established programming languages. Moreover, it has computational kernels for CPUs, GPUs, and TPUs.

\paragraph{}  Besides the methods for elementary operations such as Kronecker and Hadamard tensor multiplication, minimum location, tensor reshaping, and tensor factorizations such as Kruskal and Tucker decompositions, TensorFlow has a number of numerical optimizers which are common in deep learning such as AdaGrad. Also, TensorFlow supports checkpoints, allowing the early termination of a training process.

\paragraph{}  Within a blockchain context, TensorFlow can accelerate numerical computations for hashing or encryption. Additionally, it can be used to train a neural network, recurrent, convolutional, autoencoding, or otherwise, which can predict the volume in the immediate future, so that a bursty load of transactions can be better balanced throughout the blockchain nodes. Moreover, similar networks can be built in order to predict which blockchain user will be the next to generate a chunk of data or will ask for a transaction verification, again for load balancing purposes. Finally, large deep learning networks can in theory be deployed in order to mount an attack on the encryption protocol used by a given blockchain, but to the best of the knowledge of the authors, no such use has been recorded.

\section{Applications}\label{sec:app}

\paragraph{}  The blockchain as a ledger structure, either public or private, because of its secure and distributed design is a place for storing sensitive data such as health condition and financial transactions. Additionally, as stated earlier any intermediaries are eliminated, at least in a higher level. Thus, any fees and premiums such as taxes or bank processing fees are also, in theory at least, automatically gone.  

\paragraph{}  Regarding the digital health world, blockchain-based applications have an enormous potential. The following list contains some of the most prominent ones.
\begin{itemize}

\item  The medical records of a netizen can be stored with safety in a blockchain and can be recovered only by the certified health professional who cure the netizen regardless of their location or whether they have cured her before.

\item  Blockchain can facilitate automated monitoring of selected biomarkers by smartphones and the measurements can be compared against personalized baselines.

\item  Netizens have much improved control over their personal records and their consent can be obtained under more transparent and clear conditions.

\item  Netizens can use micropayments or mobile payments in order to procure medicines, further protecting their privacy.

\end{itemize}

\paragraph{}  Concerning the growing insurance market, there is also a significant room for blo\-ck\-chain-ba\-sed applications. Some indicative are the following:
\begin{itemize}

\item  Netizens can search easier for attractive offers and can contact insurance agents directly in order to negotiate for even better offers. This can also be done through software agents configured to look specific offers or terms.

\item  Netizens and insurance agents can hold smart contracts such as property and vehicle electronic contract purchses in blockchains. At a later point, should the need arise, they can directly renegotiate contract terms which will be also recorded in the blockchain, provided the interested parties reach an agreement.

\item  Once smart contracts are recorded, ordinary shallow or deep learning algorithms can be run atop the blockchain in order to identify possible fraud cases.  

\item  Blockchains simplify considerably payments and can even be combined with mobile payments. Payment records remain immutable and constitute proof that a payment indeed took place at the time indicated.  

\item  Claims can be automatically verified by smartphones and other personal devices which are connected to the blockchain, reducing thus the administrative burden and the overhead.

\end{itemize}

\paragraph{}  At this point it should also be reminded that the general advantages of section \ref{sec:blockchain} also hold in addition to those listed above.

\section{Conclusions}\label{sec:more}

\paragraph{}  The twofold epicenter of this \bmwork{} was the blockchain applications in the domains of digital health and insurance market and the ways Google TensorFlow, a low level computational framework for computationally intensive applications, can be used to accelerate the associated computations. The blockchain has numerous applications in the domains of medical healthcare and insurance. Moreover, it reinforces the privacy and transparency conditions and, thus, help establish a viable and scalable market.

\paragraph{}  Further research directions include the development of extensively tested blockchain management systems so that most, if not all, zero day exploits are eliminated. Moreover, given the recent advances in quantum computing which make large scale brute force attacks feasible, stronger cryptographic schemes should be sought in order to protect the sensitive personal data stored in blockchains.

\begin{acknowledgements}
We gratefully acknowledge the support of NVIDIA Corporation with the donation of the Titan Xp GPU used for this research.
\end{acknowledgements}

\bibliographystyle{spbasic}
\bibliography{genedis_2018_blockchain}

\begin{thebibliography}{24}
\providecommand{\natexlab}[1]{#1}
\providecommand{\url}[1]{{#1}}
\providecommand{\urlprefix}{URL }
\expandafter\ifx\csname urlstyle\endcsname\relax
  \providecommand{\doi}[1]{DOI~\discretionary{}{}{}#1}\else
  \providecommand{\doi}{DOI~\discretionary{}{}{}\begingroup
  \urlstyle{rm}\Url}\fi
\providecommand{\eprint}[2][]{\url{#2}}

\bibitem[{Abadi(2016)}]{abadi:2016:sigplan}
Abadi M (2016) {T}ensor{F}low: {L}earning functions at scale. {ACM} {SIGPLAN}
  {N}otices 51(9):1--1

\bibitem[{Abadi et~al.(2016)}]{abadi:2016:osdi}
Abadi M, et~al. (2016) {T}ensor{F}low: {A} system for large-scale machine
  learning. In: {OSDI}, vol~16, pp 265--283

\bibitem[{Abbas et~al.(2018)Abbas, Zhang, Taherkordi, and Skeie}]{abbas:2018}
Abbas N, Zhang Y, Taherkordi A, Skeie T (2018) {M}obile edge computing: {A}
  survey. {IEEE} {I}nternet of {T}hings {J}ournal 5(1):450--465

\bibitem[{Antonopoulos(2014)}]{antonopoulos:2014}
Antonopoulos AM (2014) {M}astering {B}itcoin: {U}nlocking digital
  cryptocurrencies. {O'R}eilly {M}edia, {I}nc.

\bibitem[{Antonopoulos(2017)}]{antonopoulos:2017}
Antonopoulos AM (2017) {M}astering {B}itcoin: {P}rogramming the open
  blockchain. {O'R}eilly {M}edia, {I}nc.

\bibitem[{Barber et~al.(2012)Barber, Boyen, Shi, and Uzun}]{barber:2012}
Barber S, Boyen X, Shi E, Uzun E (2012) {B}itter to better - {H}ow to make
  {B}itcoin a better currency. In: {I}nternational Conference on Financial
  Cryptography and Data Security, {S}pringer, pp 399--414

\bibitem[{B{\"o}hme et~al.(2015)B{\"o}hme, Christin, Edelman, and
  Moore}]{bohme:2015}
B{\"o}hme R, Christin N, Edelman B, Moore T (2015) {B}itcoin: {E}conomics,
  technology, and governance. {J}ournal of {E}conomic {P}erspectives
  29(2):213--38

\bibitem[{Cachin(2016)}]{cachin:2016}
Cachin C (2016) {A}rchitecture of the hyperledger blockchain fabric. In:
  {W}orkshop on Distributed Cryptocurrencies and Consensus Ledgers, vol 310

\bibitem[{Drakopoulos et~al.(2016)Drakopoulos, Gourgaris, Kanavos, and
  Makris}]{drakop:2016:ijait}
Drakopoulos G, Gourgaris P, Kanavos A, Makris C (2016) {A} fuzzy graph
  framework for initializing {k-m}eans. {IJAIT} 25(6):1--21

\bibitem[{Drakopoulos et~al.(2017)Drakopoulos, Kanavos, Mylonas, and
  Sioutas}]{drakop:2017:snam}
Drakopoulos G, Kanavos A, Mylonas P, Sioutas S (2017) {D}efining and evaluating
  {T}witter influence metrics: {A} higher order approach in {N}eo4j. {SNAM}
  71(1)

\bibitem[{Drakopoulos et~al.(2018{\natexlab{a}})Drakopoulos, Gourgaris, and
  Kanavos}]{drakop:2018:evos:neo4j}
Drakopoulos G, Gourgaris P, Kanavos A (2018{\natexlab{a}}) {G}raph communities
  in {N}eo4j: {F}our algorithms at work. {E}volving {S}ystems
  \doi{10.1007/s12530-018-9244-x}

\bibitem[{Drakopoulos et~al.(2018{\natexlab{b}})Drakopoulos, Liapakis, Tzimas,
  and Mylonas}]{drakop:2018:ictai}
Drakopoulos G, Liapakis X, Tzimas G, Mylonas P (2018{\natexlab{b}}) {A} graph
  resilience metric based on paths: {H}igher order analytics with {GPU}. In:
  {ICTAI}, {IEEE}

\bibitem[{Drakopoulos et~al.(2019)Drakopoulos, Stathopoulou, Kanavos,
  Paraskevas, Tzimas, Mylonas, and Iliadis}]{drakop:2019:evos:tensor}
Drakopoulos G, Stathopoulou F, Kanavos A, Paraskevas M, Tzimas G, Mylonas P,
  Iliadis L (2019) {A} genetic algorithm for spatiosocial tensor clustering:
  {E}xploiting {T}ensor{F}low potential. {E}volving {S}ystems
  \doi{10.1007/s12530-019-09267-8}

\bibitem[{Kanavos et~al.(2017)Kanavos, Drakopoulos, and
  Tsakalidis}]{drakop:2017:webist}
Kanavos A, Drakopoulos G, Tsakalidis A (2017) {G}raph community discovery
  algorithms in {N}eo4j with a regularization-based evaluation metric. In:
  {WEBIST}

\bibitem[{Kosba et~al.(2016)Kosba, Miller, Shi, Wen, and
  Papamanthou}]{kosba:2016}
Kosba A, Miller A, Shi E, Wen Z, Papamanthou C (2016) {H}awk: {T}he blockchain
  model of cryptography and privacy-preserving smart contracts. In: {IEEE}
  symposium on security and privacy, {IEEE}, pp 839--858

\bibitem[{Matthews et~al.(2017)}]{matthews:2017}
Matthews DG, et~al. (2017) {GP}flow: {A} {G}aussian process library using
  {T}ensor{F}low. {T}he {J}ournal of {M}achine {L}earning {R}esearch
  18(1):1299--1304

\bibitem[{Miller(2018)}]{miller:2018}
Miller D (2018) {B}lockchain and the {I}nternet of {T}hings in the industrial
  sector. {IT} {P}rofessional 20(3):15--18

\bibitem[{Nakamoto(2008)}]{nakamoto:2008}
Nakamoto S (2008) {B}itcoin: {A} peer-to-peer electronic cash system

\bibitem[{Pass et~al.(2017)Pass, Seeman, and Shelat}]{pass:2017}
Pass R, Seeman L, Shelat A (2017) {A}nalysis of the blockchain protocol in
  asynchronous networks. In: {A}nnual International Conference on the Theory
  and Applications of Cryptographic Techniques, {S}pringer, pp 643--673

\bibitem[{Puthal et~al.(2018)Puthal, Malik, Mohanty, Kougianos, and
  Yang}]{puthal:2018}
Puthal D, Malik N, Mohanty SP, Kougianos E, Yang C (2018) {T}he blockchain as a
  decentralized security framework. {IEEE} {C}onsumer {E}lectronics {M}agazine
  7(2):18--21

\bibitem[{Swan(2015)}]{swan:2015}
Swan M (2015) {B}lockchain: {B}lueprint for a new economy. {O'R}eilly {M}edia,
  {I}nc.

\bibitem[{Underwood(2016)}]{underwood:2016}
Underwood S (2016) {B}lockchain beyond {B}itcoin. {C}ommunications of the {ACM}
  59(11):15--17

\bibitem[{Wongsuphasawat et~al.(2018)}]{wongsuphasawat:2018}
Wongsuphasawat K, et~al. (2018) {V}isualizing dataflow graphs of deep learning
  models in {T}ensor{F}low. {T}ransactions on visualization and computer
  graphics 24(1):1--12

\bibitem[{Zyskind et~al.(2015)Zyskind, Nathan et~al.}]{zyskind:2015}
Zyskind G, Nathan O, et~al. (2015) {D}ecentralizing privacy: {U}sing blockchain
  to protect personal data. In: {SPW}, {IEEE}, pp 180--184

\end{thebibliography}
\end{document}